\documentclass[fleqn,usenatbib]{mnras}

\usepackage{newtxtext,newtxmath}

\usepackage[T1]{fontenc}

\DeclareRobustCommand{\VAN}[3]{#2}
\let\VANthebibliography\thebibliography
\def\thebibliography{\DeclareRobustCommand{\VAN}[3]{##3}\VANthebibliography}

\usepackage{graphicx}	
\usepackage{amsmath}	
\usepackage{microtype}
\usepackage{subfig}
\usepackage{comment} 

\newcommand{\sectionref}[1]{{Section~\ref{#1}}}
\newcommand{\figureref}[1]{Fig.~\ref{#1}}
\newcommand{\tableref}[1]{Table~\ref{#1}}
\defcitealias{Ransome_2021}{RHD21}

\title[H$\alpha$ environments of SNe\,IIn within $z<0.02$]{An H$\boldsymbol{\alpha}$ survey of the host environments of 77 type IIn supernovae within $\boldsymbol{z<0.02}$}

\author[C.~L. Ransome et al.]{
C.~L. Ransome,$^{1,}$ $^{2}$\thanks{E-mail: C.Ransome@2018.ljmu.ac.uk}
S.~M. Habergham-Mawson$^{1}$,
M.~J. Darnley$^{1}$,
P.~A. James$^{1}$
\newauthor and S.~M. Percival$^{1}$
\\
$^{1}$Astrophysics Research Institute, Liverpool John Moores University, Liverpool Science Park iC2, Liverpool, Merseyside, L3 5RF, UK\\
$^{2}$Isaac Newton Group of Telescopes, Edificio Mayantigo, Calle Alvarez Abreu, 70, 2 piso, E-38700 Santa Cruz de La Palma, Canary Islands, Spain\\
}

\date{Accepted XXX. Received YYY; in original form ZZZ}

\pubyear{2022}

\begin{document}
\label{firstpage}
\pagerange{\pageref{firstpage}--\pageref{lastpage}}
\maketitle

\begin{abstract}
Type IIn supernovae (SNe\,IIn) are an uncommon and highly heterogeneous class of SN where the SN ejecta interact with pre-existing circumstellar media (CSM). Previous studies have found a mass ladder in terms of the association of the SN location with H$\alpha$ emission and the progenitor masses of SN classes. In this paper, we present the largest environmental study of SNe\,IIn. We analyse the H$\alpha$ environments of 77 type IIn supernovae using continuum subtracted H$\alpha$ images. We use the pixel statistics technique, normalised cumulative ranking (NCR), to associate SN pixels with H$\alpha$ emission. We find that our 77 SNe\,IIn do not follow the H$\alpha$ emission. This is not consistent with the proposed progenitors of SNe\,IIn, luminous blue variables (LBVs) as LBVs are high mass stars that undergo dramatic episodic mass loss. However, a subset of the NCR values follow the H$\alpha$ emission, suggesting a population of high mass progenitors. This suggests there may be multiple progenitor paths with $\sim$60\% having non-zero NCR values with a distribution consistent with high mass progenitors such as LBVs and $\sim$40\% of these SNe not being associated with H$\alpha$ emission. We discuss the possible progenitor routes of SNe\,IIn, especially for the zero NCR value population. We also investigate the radial distribution of the SNe in their hosts in terms of H$\alpha$ and $r'$-band flux.
\end{abstract}




\section{Introduction}

\subsection{Type IIn supernovae}
\label{sec:intro}

First categorised by \citet[][as Seyfert I SNe]{Filippenko_1989} and \citet{Schlegel_1990}, type IIn supernovae (SNe\,IIn) account for around 7\% of the total SN population \citep{Li_2011}. SNe\,IIn are generally spectroscopically characterised by a narrow feature on the Balmer series (most obvious on the H$\alpha$ line) with full width half-maximum (FWHM) $\sim10^{2}$\,km\,s$^{-1}$ \citep{Filippenko_1997}. This narrow feature is superimposed on an intermediate width (FWHM $\sim10^{3}$\,km\,s$^{-1}$) and/or a broad component (FWHM $\sim10^{3-4}$\,km\,s$^{-1}$). These characteristic narrow features originate from the SN ejecta shocking pre-existing, dense, cold, and slow circumstellar medium (CSM) with CSM densities $\sim$\, 10$^{-13}$--10$^{-15}$\,gcm$^{-3}$ \citep{Yaron_2017}. A H$\alpha$ excess is created when emission from this interaction ionises the surrounding, unshocked, CSM and then recombines \citep{Chugai_1991, Chugai_2004}. The broader components originate from the SN ejecta. The intermediate components may originate from interaction with a massive and dense, clumpy wind \cite{Chugai_2004}, possibly further broadened by Thompson scattering \citep{Chugai_2001, Dessart_2009, Humphreys_2012, Huang_2018}. SNe\,IIn often lack the P-Cygni features seen in other SN classes \citep[with exceptions such as SN\,2012ab][]{Gangopadhyay_2020} and have a blue continuum which may originate from the strong CSM interaction \citep[][]{Stathakis_1991, Turrato_1993}. 

The CSM surrounding a SN\,IIn progenitor comes from the progenitor itself or, in some cases, a companion star. The progenitor may experience mass loss episodes toward the end of its life \citep{Smith_2014}. Using light curve modelling, \citet{Moriya_2014} found that the mass loss rate of SN\,IIn progenitors may exceed $10^{-3}$\,M$_\odot$\,yr$^{-1}$ in the decades preceding the SN explosion. In their final moments, 
the mass loss may be more eruptive and dramatic \citep{Ofek_2014, Strotjohann_2020}. This is apparent in transients such as SN\,2009ip \citep{Mauerhan_2012, Smith_2014_2009ip} where the initial eruptions were non-terminal explosions sometimes known as a SN impostor. However, in the case of SN\,2009ip, the nature of the final 2012 eruption is debated. \citet{Pastorello_2013} found that the spectrum of SN\,2009ip showed high ejecta velocities ($\sim$\,13,000\,kms$^{-1}$) between the 2009 eruption and the possible terminal 2012 events, showing that these velocities do not preclude a CCSN. Instead, \citeauthor{Pastorello_2013} suggest that the spectrum of the September 2012 eruption was similar to the September 2011 and August 2012 eruptions. Those authors suggest that SN\,2009ip may have undergone a non-terminal pair-instability event. Other non-terminal scenarios for SN\,2009ip are described by \citet{Fraser_2013} and \citet{Fraser_2015}, and \citet{EliasRosa_2018} found that the SN impostor, SNhunt\,151 was remarkably similar to SN\,2009ip, those authors suggest that may be an $\eta$\,Car-like LBV ``great'' eruption that occur in a dense CSM. However subsequent, late-time observations suggest that the final eruption of SN\,2009ip was terminal \citep[e.g.][]{Graham_2017}. Furthermore, SN\,2011fh similarly underwent SN\,2009ip-like mass loss events prior to its death as a SN and may have had a massive progenitor based on the parent cluster age of $\sim$\,4.5\,Myr \citep{Pessi_2021}. Some of these impostor events may originate from the eruptive mass loss episodes from luminous blue variable (LBV) progenitor \citep[][]{Mauerhan13a, Ofek_2014, Pastorello_2017}. CSM may also be created by super asymptotic giant branch (AGB) progenitors or companions through massive winds.

SNe\,IIn are a heterogeneous SN class. Some SNe\,IIn are generally more luminous, reaching superluminous luminosities \citep[e.g., SN\,2006gy with an absolute magnitude of --22, ][]{Ofek_2007, Smith_2007}. Some SNe\,IIn are also very long lived, remaining bright for years post-explosion \citep[e.g., SN\,1988Z, SN\,2005ip, SN\,2010jl and SN\,2015da;][]{Stathakis_1991, Turrato_1993, Filippenko_1997, Stritzinger_2012, Smith_2016b, Smith17, Tartaglia20}. Therefore, SNe\,IIn display a wide locus over the timescale-luminosity phase space of exploding transients \citep{kas11}. Some SNe\,IIn, unlike their superluminous counterparts, inhabit the standard core-collapse SN (CCSN) phase-space with absolute magnitudes between --17 and --19 \citep{Li_2011, Kiewe_2012, tadd13}. Additionally, \citet{Nyholm_2020} reported a  SN\,IIn light curve rise time bimodality; with a population of slow risers, and a separate population of fast risers. Such diversity may be explained by considering that SNe\,IIn may have multiple progenitor paths.

\subsection{Progenitor systems of type IIn supernovae}\label{sec:prog}
\subsubsection{Massive progenitors} \label{sec:massprog}

 LBVs may be considered a transitional phase in the life of a massive star where an O-type star sheds its outer H-rich layer and becomes a Wolf-Rayet (WR) star \citep{HD_1994,Weis_2020}. LBVs may expel material through massive winds and more dramatic episodic eruptions. The mass loss rates from eruptive episodes, such as the great eruption of $\eta$\,Car in the 19th century, can be as high as 1\,M$_\odot$\,yr$^{-1}$, which can lead to ejected envelopes with masses as high as 10\,M$_\odot$ \citep{Smith_2003, Smith_2010}.  LBVs are observed progenitors of at least one SN\,IIn, SN\,2005gl in NGC\,266 \citep[][]{Gal-Yam_2007}.

While LBVs can have masses in excess of 50\,M$_{\odot}$, \citet{Groh_2013} found that LBVs may arise from more intermediate mass stars with masses between $20-25$\,M$_{\odot}$. \citeauthor{Groh_2013} report that after a red supergiant (RSG) phase, the pre-SN spectra in their simulations were consistent with the spectra of LBVs. \citet{Ofek_2014}, using SNe\,IIn from the Palomar Transient Survey \citep{Law_2009, Rau_2009} found that half of SNe\,IIn in the sample had precursor events resulting in an increase in brightness prior to the SN explosion that the authors interpret as outbursts. Indeed, the progenitor of SN\,2010mc suffered large eruptive mass loss only 40 days before the SN explosion \citep{Ofek_2013}. 

There remains a question as to whether LBVs can directly end their lives as SNe. In contrast to \citet{Groh_2013}, \citet{Beasor_2020} found that  the mass loss rate of RSGs in clusters in the $20-25$\,M$_\odot$ range was not sufficient to evolve the star to an LBV. Furthermore, the stellar evolution models of \citet{Maeder_2008} could not explode a star while in the LBV phase. \citet{Dwarkadas_2011} argues that stellar models require the outer H-rich envelope of the LBV to be expelled (thus becoming a WR star) before a SN explosion can occur. They suggest that the dense CSM may have been formed over time and the LBV had evolved into a WR star before explosion. However, \citet{Groh_2013b} found that if a rotational component was added to stellar evolution models, one could explode a LBV as a SN.

Another possible progenitor of SNe\,IIn are yellow hyper-giants (YHG), which are post-RSG stars with $M_{\rm ZAMS} \approx 20-60$\,M$_{\odot}$ \citep{deJager_1998}. YHGs undergo the considerable mass loss required to form enough CSM for the SN\,IIn phenomenon and \citet{BrennanA_2021, BrennanB_2021} found that the SN\,IIn candidate AT\,2016jbu may have a YHG progenitor based on archival data.

\subsubsection{Intermediate mass progenitors} \label{sec:lowmassprog}

Some SNe\,IIn may have lower mass progenitors and alternative explosion mechanisms to the core-collapse scenario.

RSGs are the progenitors of ``normal'' SNe, such as SNe\,II-Plateau (SNe\,IIP) that are named as such due to a plateau in their light curve, due to a recombination wave in a massive H-rich envelope. RSGs suffer mass loss through massive winds and can lose a lot of their H-rich envelope, which may result in a SN\,II-Linear (SN\,IIL, named for the linear light curve decay). However, some SNe\,IIn may have RSG progenitors. \citet{Smith_2009_RSGprog} present a study on the galactic RSGs, Betelgeuse and VY\,CMa. Those authors found that while Betelgeuse had steady mass loss via winds, VY\,CMa suffered more episodic mass loss. \citeauthor{Smith_2009_RSGprog} conclude that when considering the density of the CSM surrounding VY\,CMa, this RSG could be the progenitor for a SN\,1988Z-like SN\,IIn. Another SN\,IIn with a possible RSG progenitor is SN\,1998S where the CSM was produced by a strong wind and also is very dusty \citep{Meikle_2003, Mauerhan_2012, tad15}. 

Electron-capture SNe (ecSNe) arise when a star at the very lowest mass range for CCSNe ($8-10$\,M$_{\odot}$) explodes via  collapse of its ONeMg degenerate core, resulting from electron capture by $^{24}$Mg and $^{20}$Ne. Compared to typical SNe\,II explosions, ecSNe are not as energetic \citep{Miyaji_1980, Nomoto_1984, Nomoto_1987}. Prior to explosion, the progenitors of ecSNe are super-AGB (sAGB) stars and have a CSM cocoon formed from massive winds that creates the SN\,IIn spectral features.  There may be an example of a Galactic ecSN in SN\,1054 when one considers the remnant, the Crab Nebula \citep[M\,1][]{Mayall_1942, Duyvendak_1942, Smith_2013,Moriya_2014}. Another clue that ecSNe may produce SNe\,IIn is that contemporary reports of SN\,1054 are not consistent with the lower luminoisity expected from ecSNe. \citet{Moriya_2014} investigated whether this higher `historical' luminosity may be explained with CSM interaction.

Alternative to the core-collapse scenario, a SN\,Ia exploding in a dense CSM could produce spectral features similar to that of a SN\,IIn \citep{Deng_2004,Dilday_2012, Silverman_2013}. 

\citet{Hamuy_2003} found that SN\,2002ic was the first unambiguous SN\,Ia-CSM and showed the classic complex, multicomponent H Balmer profiles we expect in SNe\,IIn. A number of SNe\,IIn have been identified as SNe\,Ia-CSM at later times when the characteristic SN\,Ia features become apparent such as broad \ion{Si}{ii} absorption. Examples include SN\,1997cy \citep{Germany_2000, Prieto_2005} and SN\,2005gj \citep{Prieto_2005, Aldering_2006}. The classification of SNe\,Ia-CSM can be challenging as the identifying SN\,Ia features may become apparent at later times. The recent findings of \citet{Jerkstrand_2020} suggest the well studied superluminous SN\,IIn, SN\,2006gy may be thermonuclear in origin. \citeauthor{Jerkstrand_2020} note that the neutral Fe lines seen in the spectra of SN\,2006gy at late times (+394\,days) were consistent with their SN\,Ia models where the SN ejecta hit dense CSM.

\subsubsection{Supernova Impostors } \label{sec:imps}

SN impostors are a part of a more general group known as gap transients that inhabit the gaps in the timescale-luminosity phase space between classical novae and SNe \citep{kasl11}. SN impostors are typically subluminous compared to `true' SNe but are more luminous than novae, with impostors having absolute magnitudes $M_V \approx-11$ to $-14$ \citep{Kochanek_2012}. These SN impostors may be the great outbursts of LBVs, similar to the great eruption of $\eta$\,Car \citep{Smith_2011}. The progenitor may subsequently be obscured by dust produced after the eruption \citep{Kochanek_2012}, however, this is not the case for all impostors, such as SN\,2002kg \citep{Kochanek_2012, Humphreys_2014, Humphreys_2017}.  

It is possible that SN impostors can precede a `true' SN such as SN\,2009ip in NGC\,7259 \citep{Foley_2011, Mauerhan13a, Pastorello_2013} and SN\,2015bh \citep{Boian_2018, Thone_2017}. However, in the case of SN\,2009ip, where there were two outbursts, in 2009 and 2012. \citet{Pastorello_2013} and \citet{Smith_2014_2009ip} found that the 2012 eruption was consistent with a terminal CCSN explosion, taking into consideration the high luminosity, explosion energy and the enduring emission features. \citet{Fraser_2013}, however, reported that the spectra in 2012 were similar to the 2009 eruption, the luminosity had not dropped below pre-discovery levels and there was a lack of expected nucleosynthesised elements. 

A noteworthy transient that is the prototype for its own subclass is SN\,2008S in NGC\,6964 \citep{Arbour_2008, Thompson_2009}. The progenitor could not be recovered in optical pre-explosion imaging and \citet{Prieto_2008} suggest the progenitor may have a mass of $\sim$\,10\,M$_\odot$ and inhabited a dust-rich environment that shrouded progenitor and thus the transient was an impostor. However, \citet{Botticella_2009} reported that the light-curve for SN\,2008S was consistent with being powered by $^{56}$Co decay and that the progenitor may have been an AGB star and the explosion was a terminal ecSN. \citet{Adams_2016} found that SN\,2008S had faded, by 2015 to a level dimmer than its progenitor, which may suggest the explosion was terminal, or extreme dust behaviour must be invoked. Another example of a SN impostor is SNhunt248 that had an optical peak magnitude of -15 and appeared to be a YHG with ejecta from the eruption interacting with CSM \citep{Kankare_2015}.

\subsection{Environmental studies} \label{sec:env}

The local environment of a SN within its host galaxy can offer information on the possible progenitor system. For example, one can examine the association of SNe to ongoing star-formation as traced by H$\alpha$ emission. Generally speaking, SNe\,II trace regions of recent star formation and the most massive stars may trace ongoing star formation \citep[see, for example][]{Anderson_2009, hab14}. H$\alpha$ emission indicates a characteristic time-scale of under 16\,Myr \citep{Haydon_2020}. Possible high mass and short lived progenitors, such as LBVs,  would be expected to be found in regions of ongoing star formation, which would be traced by H$\alpha$ emission. 

Another way the environments of SNe have been probed is through the radial distributions of SNe in terms of the observed flux in a particular filter (we describe this method in \sectionref{rad}). \citet{hab14} found that the radial distribution of SNe\,IIn in their sample were very different from the radial distributions of SNe\,Ic. The SNe\,Ic were centrally located (hence a higher metallicity region) and they did not find a central concentration of SNe\,IIn. This suggests the two classes arise from different stellar populations. 

 \citet{hab14} use the O3N2 diagnostic \citep{pet04} to compare the metallicity at the sites of different SN classes. It was found that SN\,IIn environments are more metal rich than SNe\,IIP and that the local metallicities of SN impostors tend to be lower than SNe\,Ic, IIn or IIP. \citet{tad15} use optical spectra of 60 interacting transients (SNe\,IIn, SNe\,Ibn and impostors). \citeauthor{tad15} found that impostors were generally in regions of lower metallicity when compared to SNe\,IIn. SNe\,IIn similar to SN\,1998S were found in higher metallicity regions than SNe\,IIL and SNe\,IIP, and long lasting SNe\,IIn, such as SN\,2005ip and SN\,1988Z, were in regions of lower metallicity as seen in the environments of SN impostors. \citeauthor{hab14} conclude that SN\,IIn environment variation suggests there may be multiple progenitor paths rather than solely LBVs.

\subsection{This paper}
In this paper, we use a vastly expanded SN\,IIn sample to probe the possible progenitor paths using local host galaxy information . Using pixel statistics techniques and observations from our H$\alpha$ SNe\,IIn host survey, we present the largest environmental study of SNe\,IIn to date. We outline our SN\,IIn sample selection, observations and data analysis in \sectionref{methods}. In \sectionref{results} we present our target catalogue and results of our host galaxy analysis along with example images of the hosts. We then discuss our results and their implications for the progenitor channels of SNe\,IIn in \sectionref{Discuss}.

\section{Methods} \label{methods}

\subsection{The classification of type IIn supernovae and our sample} \label{sec:class}

Perhaps due to the diversity of SNe\,IIn, classification  can be difficult. One may group SNe\,IIn into groups of similar objects, for example,  SN\,1988Z-like which have slow photometric evolution, SN\,1994W-like with a plateau in the light curve lasting a few months preceding a rapid decline, and SN\,1998S-like that have a fast rise and decline time in the light-curve. \citet[][ hereafter \citetalias{Ransome_2021}]{Ransome_2021} introduced a classification scheme for SNe\,IIn, mostly based on the complex, multi-component H$\alpha$ profiles. In this scheme, transients are split into three groups, gold, silver, and not SNe\,IIn. Gold SNe\,IIn exhibit enduring CSM interaction features, silver SNe\,IIn show weaker CSM interaction or have a single spectrum available. \citetalias{Ransome_2021} found that 28 of SNe of the 87 in their sample of SNe, may have previously been misclassified. \citetalias{Ransome_2021} produced a robust, systematically classified sample of SNe\,IIn which we use in our environmental analysis. 

Our sample is outlined in \tableref{fulltab}. The table contains the SNe, \citetalias{Ransome_2021} spectral category and the telescope used. Most of our observations are in the Northern hemisphere, however, three of our SNe\,IIn observed with LCOGT\,2m located in Australia.

\subsection{Observations}

In the Northern Hemisphere, we utilised the Liverpool Telescope \citep[LT;][]{Steele_2009} and Isaac Newton Telescope (INT) at the Observatorio de Roque de las Muchachos on La Palma in the Canary Islands. For our observations on the LT, we use the IO:O instrument\footnote{https://telescope.livjm.ac.uk/TelInst/Inst/IOO/}.  We use the Wide Field Camera (WFC) on the INT\footnote{https://www.ing.iac.es//Astronomy/instruments/wfc/}. In the Southern Hemisphere, we use the Las Cumbres Observatory 2m (LCOGT\,2m) at Siding Springs Observatory, New South Wales, Australia \citep{lco}. We use Spectral\footnote{https://lco.global/observatory/instruments/spectral/} on the LCOGT\,2m.

Our observations consist of 3\,$\times$\,300\,s exposures using the appropriate redshifted H$\alpha$ filter and a single 300\,s exposure with in the $r'$-band with 2\,$\times$\,2 binning. These observations span 2019 to 2021. We limit our observations to nearby (z\,<\,0.02) hosts as we require that \ion{H}{ii} regions are resolved for the pixel statistics described in \sectionref{NCR}. Furthermore, we select hosts with axial ratios of under 4:1 to ensure the SN\,IIn site is associated with the observed emission, rather than possibly coincident as may be the case in a more edge-on galaxy. Major and minor axes for each host were taken from the SIMBAD Astronomical Database \citep{simbad}. Both IO:O and Spectral have a $10^\prime$ field of view and all surveyed hosts were contained within the field of view. The LT has redshifted H$\alpha$ filters covering the range to around z\,=\,0.04 and all of our targets are within z\,=\,0.02. We only use the LCOGT\,2m for nearby targets as there is only a rest-frame H$\alpha$ filter. 

Raw data from the LT and LCOGT\,2m are reduced by the respective standard pipelines \citep{ioo, lco}. The NCR analysis requires continuum-subtracted images. We use the methods in \citet{and12} but using IRAF \citep{IRAF}, astropy \citep{astropy} and specutils \citep{specutils}. Data from the INT was reduced using standard procedures using astropy. 

\subsection{Pixel statistics: normalised cumulative ranking}\label{NCR}

NCR was first implemented by \citet{ja06} and subsequently utilised in environmental studies \citep[e.g.,][]{and08, and12, hab14}. NCR traces the association of a pixel with some emission, in our case, to H$\alpha$ emission and therefore, the association to ongoing star-formation. NCR assigns every pixel a value between zero and one. NCR is calculated by ordering the the pixel values of the continuum-subtracted H$\alpha$ image in ascending order, summing these values and normalising by the sum. A zero NCR value indicates no H$\alpha$ emission and therefore no association with ongoing star-formation. An NCR value of one indicates the strongest star-forming pixel within the entire galaxy image. After continuum subtraction, some pixels will have negative values. Pixels before the sum turns positive are assigned an NCR value of zero. The NCR value of a pixel may be written as,

\[\mathrm{NCR_{i}} = \frac{\sum_{1}^{i}x_{i}}{\sum_{1}^{N}x_{N}},\]

\noindent where $i$ is the  current pixel, $N$ is the total number of pixels and $x$ is the pixel value.

Using NCR analysis, \citet{and08} found that SNe\,IIn in their sample did not follow the H$\alpha$ emission in their hosts. Those authors did, however, note that the SNe\,IIn more closely followed near-ultraviolet (NUV) emission when comparing data from the \textit{Galaxy Evolution Explorer} \citep[GALEX,][]{galex}. NUV traces recent star-formation with a characteristic time-scale of over around 19.6\,Myr for \textit{GALEX} NUV \citep{Haydon_2020}. 

\citet{and12} set out to constrain progenitor properties using the SN association to H$\alpha$ and NUV emission. They found that there was a mass ladder in terms of association to H$\alpha$ emission with the most massive progenitors being most associated with ongoing star formation, owing to their shorter lifetimes. It was found that the mass sequence, starting with SNe\,Ic \citep[see also,][]{Kangas_2013} as the most massive and most associated, went as such:

\[\mathrm{SNe\,Ic}\,\rightarrow\,\mathrm{SNe\,Ib}\,\rightarrow\,\mathrm{SNe\,II}\,\rightarrow\,\mathrm{SNe\,Ia}.\]

\noindent \citeauthor{and12} suggested that this indicated that the majority of SNe\,IIn do not have high-mass progenitors, such as LBVs. However, \citet{snt15} noted that LBVs are often not associated with clusters of O-stars as one would expect of a massive progenitor. Instead those authors suggest that many LBVs may form in binary systems where the LBV is a mass gainer and a WR star is a donor in situ in the home O-star cluster. 

\citet{hab14} investigated the environments of SNe\,IIn and SN impostors. There were 24 SNe\,IIn in their sample and again it was found that these transients did not follow star formation as traced by H$\alpha$, via the NCR method. \citeauthor{hab14} conclude that as some SNe\,IIn do have LBV progenitors, there may be multiple progenitor channels for SNe\,IIn.

Additionally, \citet{Kangas_2017} utilised the NCR method with resolved massive stars in the Large Magellanic Cloud (LMC) and M33. \citeauthor{Kangas_2017} compare the NCR distributions of the massive stars in their sample with the NCR distributions of the different classes of SNe in \citet{and12}. \citeauthor{Kangas_2017} note that SNe\,IIn are a diverse class, possibly with multiple progenitors. Those authors suggest that when combining the NCR values of RSGs (M\,$\lesssim$\,8\,M$_\odot$) and LBVs (with RSGs making up 70\% and LBVs accounting for 30\% of the NCR values), the average SN\,IIn NCR value can be reproduced.

To compute the NCR value, the average NCR value of a 3\,$\times$\,3 pixel bin centred around the target pixel is taken. We apply this to our sample and then make cumulative distributions. We split our sample into the spectral categories in \citetalias{Ransome_2021} and then compare these distributions to each other and also a hypothetical 1:1 relation which represents an NCR distribution that `perfectly' follows the emission.

We can then implement Anderson-Darling \citep[AD,][]{ADTest, ADTest2} and Hartigan dip tests \citep[HDT][]{hartigan}, as well as the bimodality coefficient \citep{bimodcoeff} in order to test whether there may be multiple populations of SNe\,IIn in terms of their association to ongoing star-formation as traced by H$\alpha$ emission.

\subsection{Radial analysis} \label{rad}

The radial distribution of SNe can indicate the general stellar population. In spiral galaxies, the bulge tends to have an older population while the discs have younger populations with active star formation. Furthermore, the spatial distribution within a galaxy may provide an analogue for metallicity as there is a metallicity gradient \citep{Henry_1999} with central regions having a higher metallicity than further out in the disc regions.

\begin{figure} 
  \includegraphics[width = \columnwidth]{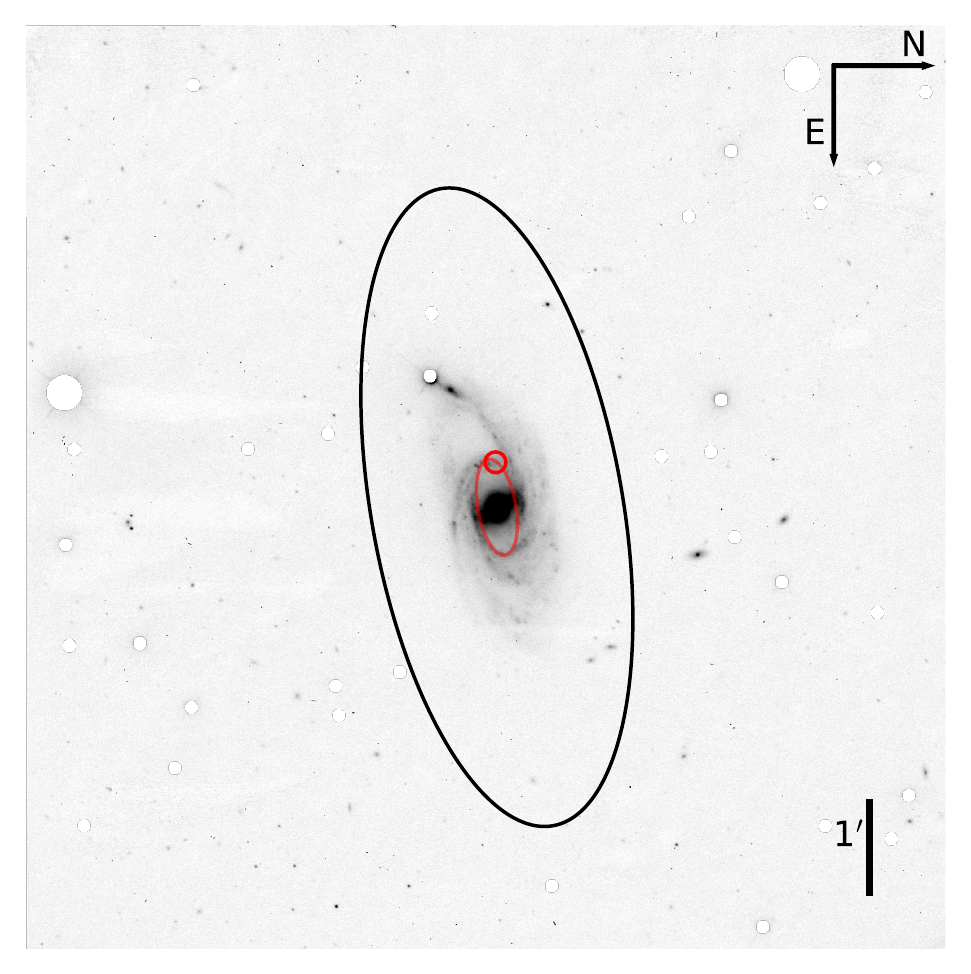}
  \caption{An example of the radial analysis employed in this paper. Shown is the $r'$-band image of the host of PTF\,11iqb, NGC\,151 (z=0.0126). The red ellipse just intercepts the SN\,IIn site (red circle). The black ellipse encloses the galaxy $r'$-band flux. These ellipses are also used in the continuum subtracted H$\alpha$ image. In order to measure the emission from the host, foreground stars were masked.}
  \label{iqb_rad}
\end{figure}

We measure and compare the radial distribution of our SNe\,IIn in terms of the $r'$-band and H$\alpha$ flux in their host galaxies. We adopt the method used in \citet{Anderson_2009}. This is achieved calculating the ratio of ellipses that cover the extent of the host $r'$-band light and the ellipse enclosing the SN position. Ellipse parameters were taken from the NASA/IPAC Extragalactic Database \citep[NED][]{NED}. For the ellipse covering the total $r'$-band emission from the host, the semi-major axis is increased until the ellipse has reached the sky level on the $r'$-band image. The sky level of the image is determined such that the difference in enclosed flux between ellipses has flattened out to being around zero. This ellipse is then used with the continuum subtracted H$\alpha$ image. This gives us Fr(R), the fraction of $r'$-band light contained within the SN ellipse and Fr(H$\alpha$), the fraction of H$\alpha$ emission contained in the SN ellipse. Therefore each SN will have a Fr(R) and an Fr(H$\alpha$) value of between 0 and 1, where 0 indicates the SN is at the centre of the host and a value of 1 indicates the SN is at the extreme periphery of the host. For this radial analysis, it is important to mask stars such that the measured flux is from the host. Using the results from the \texttt{starfind} and \texttt{xyxymatch} subroutines in IRAF, the (non-saturated) stars are detected and masked. Saturated stars are not detected by \texttt{starfind} so these are masked by eye. This is carried over to the continuum subtracted H$\alpha$ images to eliminate artefacts that can be left over after continuum subtraction. In \figureref{iqb_rad} we present an example of this radial analysis, where we show the ellipse enclosing the full $r'$-band flux of the host galaxy and the ellipse that just encloses the SN pixel.

Using these statistical measures, we will now analyse the SN\,IIn populations in host galaxies within z\,=\,0.02.

\section{Results} \label{results}

\subsection{The hosts}

Our observing campaign spanning from July 2019 -- July 2021 increased the sample size of SN\,IIn hosts (within z < 0.02) from previous studies (24) to 78. Of these, we observe 30 gold and 34 silver SNe\,IIn classified by \citetalias{Ransome_2021}. The remaining observations were hosts of SN\,IIn without any public spectral data so were not spectrally reclassified but are classified as SNe\,IIn on public databases.

\begin{figure*} 
\centering
  \includegraphics[width = \textwidth]{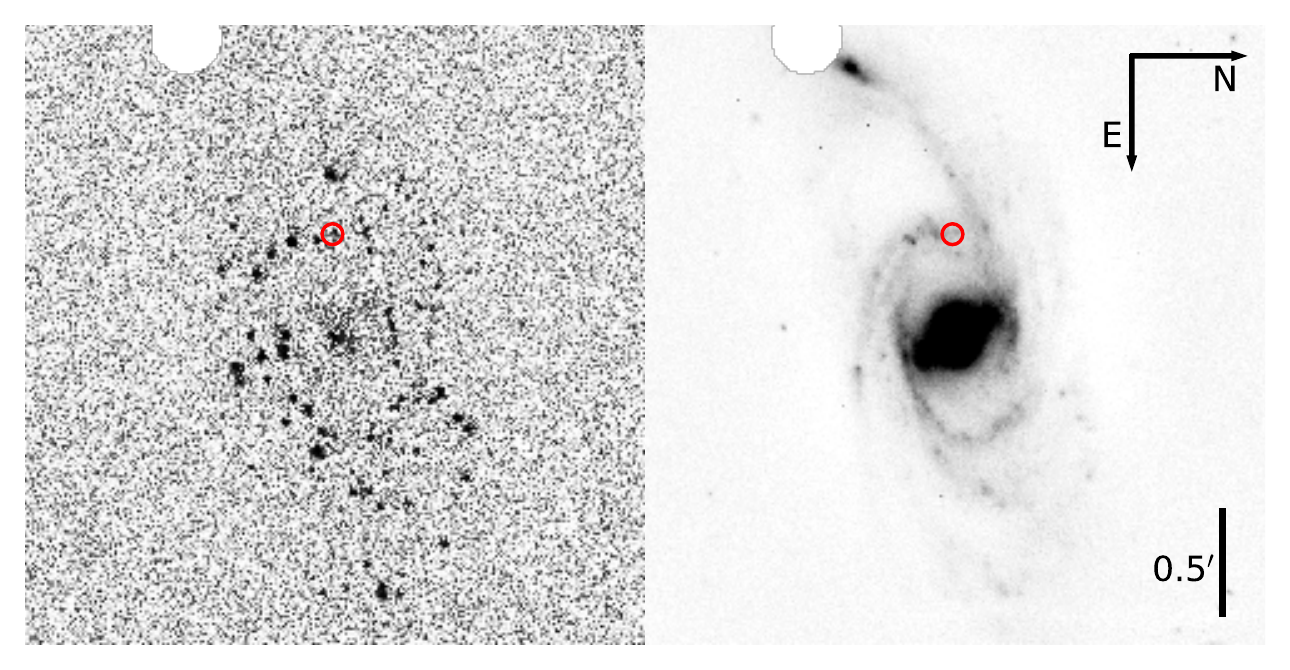}
  \caption{The continuum-subtracted H$\alpha$ (\textit{left}) and $r'$-band (\textit{right}) environments of PTF\,11iqb in NGC\,151 (z=0.0126). The position of PTF\,11iqb is marked with the red circle. PTF\,11iqb is an example of a SN\,IIn associated with star-formation as traced by the H$\alpha$ emission, resulting in an NCR value of 0.845. The foreground star at the top of the image has been masked out in the H$\alpha$ image as the continuum subtraction leaves artefacts which may interfere with the NCR value calculation.}
  \label{ptf11iqb}
\end{figure*}

\begin{figure*} 
\centering
  \includegraphics[width = \textwidth]{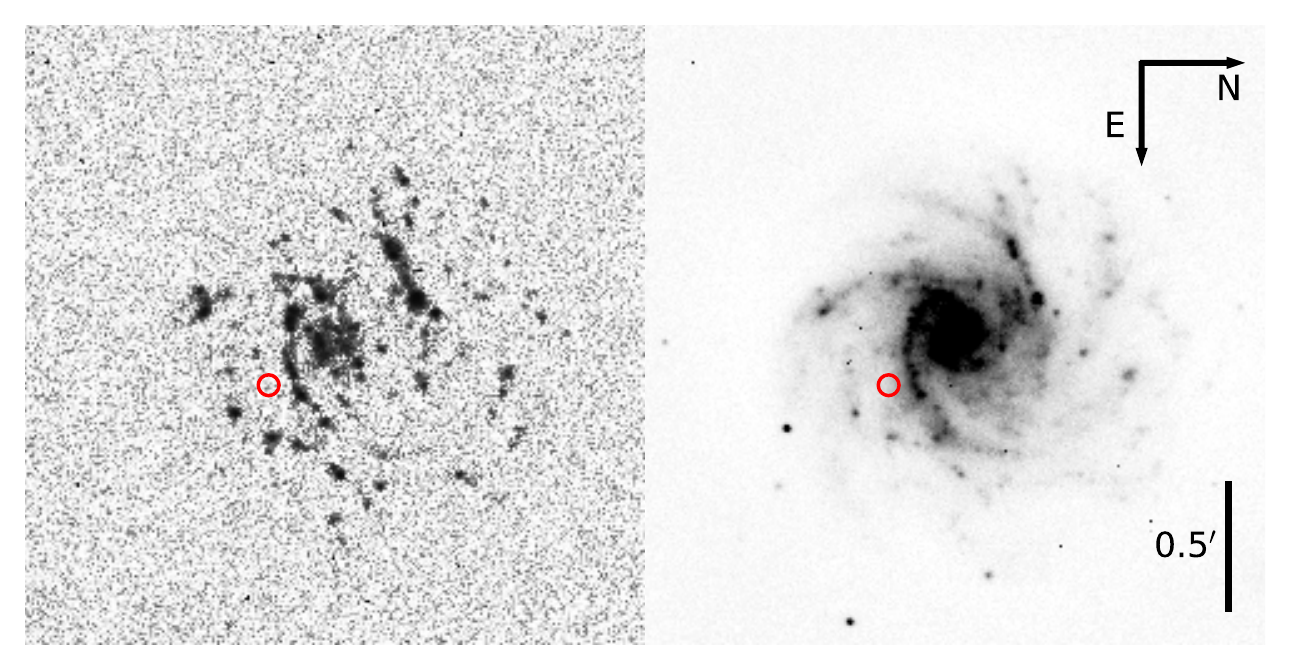}
  \caption{The continuum-subtracted H$\alpha$ (\textit{left}) and $r'$-band (\textit{right}) environments of SN\,2003lo in NGC\,1376 (z=0.0139). The position of SN\,2003lo is marked with the red circle. SN\,2003lo is an example of a SN\,IIn that is not associated with star-formation as traced by the H$\alpha$ emission (NCR value of zero). In the case of SN\,2003lo, the transient resides in an apparent inter-arm gap.}
  \label{2003lo}
\end{figure*}

In \figureref{ptf11iqb} -- \figureref{2003lo} we present the two of our SNe\,IIn shown in both the $r'$-band and continuum subtracted H$\alpha$ images. We also mark the SN positions with a red circle. We present the full sample of SNe\,IIn we have host observations for in \tableref{fulltab}.

In \tableref{fulltab} we present our full sample of 77 SNe\,IIn with their NCR values, classification from \citetalias{Ransome_2021}, Fr(H$\alpha$) values, Fr(R) values, and the telescope used in the observations. Most of our observations were from the Northern Hemisphere, however three of our SNe\,IIn are southern targets observed with LCOGT\,2m. The Southern Hemisphere sub-sample is somewhat small due to the limited H$\alpha$ filters available at LCOGT\,2m. 

\begin{table*}
\caption{Our sample of 77 SNe\,IIn and their hosts with the discovery date, coordinates, NCR value, Fr(H$\alpha$) values, Fr(R) value, the telescope we used for the observation and the classification from \citetalias{Ransome_2021}.\label{fulltab}}
\begin{center}
\begin{scriptsize}
\begin{tabular}{lllllllrlll}
\hline
\multicolumn{1}{l}{Name} & \multicolumn{1}{l}{Disc.\ Date} & \multicolumn{1}{l}{Host} & \multicolumn{1}{l}{R.A.} & \multicolumn{1}{l}{Dec.} & \multicolumn{1}{l}{z}& \multicolumn{1}{l}{Telescope} & \multicolumn{1}{l}{NCR value} & \multicolumn{1}{l}{Classification} & \multicolumn{1}{l}{Fr(H$\alpha$)} &\multicolumn{1}{l}{Fr(R)}\\
\hline
    SN\,2005db & 2005/07/19& NGC\,214& 00:41:26.79& 25:29:51.60& 0.0151 &LT&0.387&Gold & 0.528& 0.526\\
    SN\,2005gl & 2005/10/05&NGC\,266 &00:49:50.02 &32:16:56.80 &0.0155 &LT & 0.765& Gold & 0.644&0.526  \\
    SN\,1999eb & 1999/10/02 &NGC\,664 &  01:43:45.45 & 04:13:25.90& 0.0180 &LT &0.750&Gold &0.324 & 0.258 \\
    SN\,2003G &2003/08/01 & IC 208 & 02:08:28.13 & 06:23:51.90&0.0120 &LT & 0.000&Gold &- & - \\
    SN\,2008J &2008/01/15 &  MCG\,-02-07-33 & 02:34:24.20 & -10:50:38.50&0.0159 &LT &0.808 &Gold & 0.069 &0.077 \\
    SN\,2000eo &2000/11/16 & MCG -02-09-03 & 03:09:08.17 & -10:17:55.3&0.0100 &LT &0.000&Gold  & -& -\\
    SN\,2006gy &2006/09/18 & NGC 1260 & 03:17:27.06 & 41:24:19.5&0.0192 &LT &0.907&Gold  & 0.014& 0.030 \\    
    SN\,1989R &1989/10/14 & UGC 2912 & 03:59:32.56 & 42:37:09.20& 0.0180 &LT &0.199 & Gold & -& -\\
    SN\,1995G & 1995/02/23& NGC 1643 & 04:43:44.26 & -05:18:53.70& 0.0160 &LT &0.007 &Gold &0.549 & 0.620 \\
    SN\,2006jd & 2006/10/12&  UGC 4179 & 08:02:07.43 & 00:48:31.50& 0.0186 &LT &0.879 &Gold  &0.411 &0.580 \\
    SN\,2009kn & 2009/10/26& MCG\,-03-21-06 & 08:09:43.04 & -17:44:51.30&0.0143 &LT &0.000 &Gold &- &- \\
    SN\,2005kj & 2005/11/17& A084009-0536 & 08:40:09.18 & -05:36:02.20& 0.0160 &LT &0.000 &Gold &- &- \\
    SN\,1994ak & 1994/12/24& NGC 2782 & 09:14:01.47 & 40:06:21.50& 0.0085 &LT &0.000 &Gold &0.955 &0.869 \\
    SN\,2005ip & 2005/11/05& NGC 2906 & 09:32:06.42 & 08:26:44.40& 0.00718 &LT & - &Gold &0.685 &0.540\\
    SN\,1989C & 1989/02/03& MCG +01-25-25 & 09:47:45.49 & 02:37:36.10& 0.0063 &LT &0.830 &Gold &- & -\\
    SN\,2011ht & 2011/09/29& UGC 5460 & 10:08:10.56 & 51:50:57.12& 0.0036 &LT &0.000 &Gold &0.090 &0.197 \\
    SN\,1993N & 1993/04/15& UGC 5695 & 10:29:46.33 & 13:01:14.00& 0.0098 &LT &0.000 &Gold &0.485 & 0.490 \\
    SN\,1998S & 1998/03/02& NGC 3877 & 11:46:06.13 & 47:28:55.40& 0.0030 &LT &0.780 &Gold &- & -\\
    SN\,1994W &1994/07/29 &NGC 4041 & 12:02:10.92 & 62:08:32.70& 0.0040 &LT &0.000 &Gold & 0.698&0.510 \\
    SN\,2011A & 2001/01/02& NGC 4902 & 13:01:01.19 & -14:31:34.80& 0.0089 &LT &0.000 &Gold &0.385 &0.607 \\
    SN\,2016bdu & 2016/02/28& - &  13:10:13.95 & 32:31:14.07& 0.0170 &LT &0.707 &Gold &- &- \\
    SN\,1997eg & 1997/12/04& NGC 5012 & 13:11:36.73 & 22:55:29.40& 0.0087 &LT &0.784 &Gold &0.454 & 0.523 \\
    SN\,2015da &2015/01/09 & NGC 5337 & 13:52:24.11 & 39:41:28.60& 0.0072 &LT & - &Gold & 0.447 & 0.564 \\
    SN\,1994Y & 1994/08/19&  NGC 5371 & 13:55:36.90 & 40:27:53.40& 0.0085 &LT &0.000 &Gold &0.112 & 0.320 \\
    SN\,1995N & 1995/05/05& MCG-02-38-17 & 14:49:28.29 & -10:10:14.40& 0.0062 &LT &0.000 &Gold &- &- \\
    SN\,2008B & 2008/01/02& NGC 5829 & 15:02:43.65 & 23:20:07.80& 0.0188 & LT&0.000 &Gold & 0.201& 0.383\\
    SN\,1987B & 1987/02/24& NGC 5850 & 15:07:02.92 & 01:30:13.20& 0.0085 &LT &0.107 &Gold &0.998 &0.999 \\
    SN\,2008S & 2008/02/01& NGC 6946 & 20:34:45.35 & 60:05:57.80& 0.0002 & LT& 0.000&Gold &- & -\\
    SN\,1999el & 1999/10/20& NGC 6951 & 20:37:18.03 & 66:06:11.90& 0.0047 &INT &0.000 &Gold &0.277 & 0.321\\
    SN\,2009ip & 2012/07/14& NGC 7259 & 22:23:08.30 & -28:56:52.40& 0.0059 &LCOGT\,2m&0.000 &Gold &0.788 &0.891 \\
    SN\,2019el & 2019/01/02 & - & 00:02:56.70& +32:32:52.30& 0.0005 &LT &0.003&Silver  & -&- \\
    SN\,2017hcc & 2017/10/02& GALEX 2.67E+18 &00:03:50.58 &-11:28:28.78& 0.0173 &LT &0.993&Silver  &- &- \\
    SN\,2011fx &2011/08/30& MCG+04-01-48 & 00:17:59.56&24:33:46.00& 0.0193 &LT &0.726&Silver  &- &- \\
    PTF\,11iqb & 2011/07/23& NGC\,151&00:34:04.84 & -09:42:17.90& 0.0125& LT & 0.845&Silver & 0.059 & 0.334 \\
    SN\,2007pk &2007/11/10 &NGC\,579 &01:31:47.07 & 33:36:54.10& 0.0167 & LT&0.785&Silver &0.248 &0.108  \\
    SN\,2016eem &2016/07/08 & -  &02:05:59.80 & 47:44:14.00&0.0200 &LT & 0.231 & Silver & -&- \\
    SN\,2002ea &\,2002/07/21\, & NGC 820 & \,02:08:25.08\, & \,14:20:52.80\, & 0.0148&LT &0.096 &Silver &0.124 &0.308  \\
    SN\,1978K &1978/07/31 & NGC 1313 & 03:17:38.60 & -66:33:04.60& 0.0016 &LCOGT\,2m &0.705&Silver &- &-  \\
    SN\,2003lo & 2003/12/31&  NGC 1376 & 03:37:05.12 & -05:02:17.30& 0.0140 &LT &0.000 & Silver &0.288 &0.391 \\
    SN\,2005aq & 2005/03/07& NGC 1599 & 04:31:38.79 & -04:35:06.80& 0.0130 &LT &0.370 &Silver &0.399 &0.291 \\
    Gaia14ahl & 2014/09/20& PGC 1681539 & 04:42:12.09 & 23:06:15.00& 0.0170 &LT & 0.000&Silver & -&- \\
    SN\,2005ma & 2005/12/24& MCG\,-02-13-13 & 04:49:53.91 & -10:45:23.40& 0.0150 &LT &0.262 &Silver & 0.374 &0.373 \\ 
    SN\,2016hgf & 2016/10/16&  WEIN\,69 & 04:51:45.97 & 44:36:03.06& 0.0172 &LT  &0.000&Silver  &0.280 & 0.217 \\
    SN\,2019rz & 2019/01/14&  UGC 3554 & 06:50:25.80 & 43:03:11.70 & 0.0189 &LT&0.926 &Silver & 0.028 &0.001 \\
    AT\,2018lkg & 2018/12/30&  UGC 3660 & 07:06:34.76 & 63:50:56.90& 0.0142 &LT&0.880 &Silver & 0.092 & 0.001 \\
    AT\,2014eu & 2014/11/17&  MCG+09-13-02 & 07:28:55.97 & 56:11:46.20 & 0.0179 &LT &0.228 &Silver &- &- \\
    SN\,2014ee & 2014/11/12& UGC 4132 & 07:59:11.68 & 32:54:39.60& 0.0174 &LT &0.590 &Silver &0.796 &0.655 \\
    SN\,2002fj & 2002/09/12&  NGC 2642 & 08:40:45.10 & -04:07:38.50& 0.0140 &LT &0.414 &Silver & 0.268& 0.348\\
    SN\,2015bh & 2015/02/07& NGC 2770 & 09:09:34.96 & 33:07:20.40& 0.0064 & LT&0.000 &Silver & 0.601 &0.603 \\
    SN\,2014es & 2014/11/20& MCG -01-24-12 & 09:20:46.91 & -08:03:34.00& 0.0196 &LT &0.000 &Silver & 0.900& 0.876\\
    SN\,1997ab &1997/02/28 & A095100+2004 & 09:51:00.40 & 20:04:24.00& 0.0130 &LT &0.000 &Silver &- &- \\
    SN\,1996bu & 1996/1/14& NGC 3631 & 11:20:59.18 & 53:12:08.00& 0.0039 &LT &0.026 &Silver &- &-\\
    SN\,1987F & 1987/03/22 & NGC\,4615 & 12:41:38.99 & 26:04:22.40& 0.0160 &LT & 0.000&Silver & 0.132 & 0.152 \\
    SN\,2008ip & 2008/12/31& NGC 4846 & 12:57:50.20 & 36:22:33.5& 0.0151 &LT &0.000 &Silver &0.227 &0.902 \\
    SN\,2006am & 2006/02/22&NGC 5630 & 14:27:37.24 & 41:15:35.40& 0.0089 &LT &0.555 &Silver & 0.894 & 0.683 \\
    SN\,2003dv & 2003/04/22& UGC 9638 & 14:58:04.92 & 58:52:49.90& 0.0076 &LT &0.180 &Silver & 1.000 & 0.777 \\
    SN\,2016bly & 2016/04/29& 2MASX J17224883+1400584 & 17:22:48.90 & 14:00:59.88& 0.0194 &LT &0.840 &Silver & -& -\\
    SN\,2017gas & 2017/08/10& 2MASX J20171114+5812094 & 20:17:11.32 & 58:12:08.00& 0.0100 & LT&0.920 &Silver &- & -\\
    SN\,2006bo & 2006/04/05 & UGC\,11578 &20:30:41.90&	09:11:40.80& 0.0153  & LT & 0.000 & Silver &- &- \\
    SN\,2018hpb & 2018/10/25& - &22:01:34.52 & -17:27:45.22& 0.0177 &LT &0.000 &Silver &- & -\\
    SN\,2013fs & 2013/10/07& NGC 7610 & 23:19:44.70 & 10:11:05.00& 0.0119 &LT&0.345 &Silver & 0.515& 0.660\\
    SN\,2015bf & 2015/12/12& NGC 7653 & 23:24:49.03 & 15:16:52.00& 0.0142 &LT &0.657 &Silver & 0.349&0.630 \\
    SN\,2010jj &2010/11/03 &  NGC 812 & 02:06:52.23 & 44:34:17.50 & 0.0172 & LT& 0.000& - &0.103 &0.179  \\
    PS\,15cwt &2015/08/20 & -& 02:33:16.24 & 19:15:25.20& 0.0135 & LT& 0.012& - &- & -\\
    SN\,2011js & 2011/12/31 &NGC\,1103 & 02:48:04.96 & -13:57:51.10& 0.0138 & LT& 0.000&-  & 0.262& 0.385\\
    SN\,2006qt & 2006/10/11& A034002-0434 & 03:40:02.72 & -04:34:18.70& 0.0100 &LT &0.000& -  & -& -\\
    SN\,2005kd &2005/11/12 & PGC 14370 & 04:03:16.88 & 71:43:18.90& 0.0150 &LT &0.000 &- &- &- \\
    SN\,2007ak & 2007/03/10& UGC\,3293 & 05:20:40.75 & 08:48:16.00& 0.0156 &LT &0.000 &- &0.184 &0.130 \\
    SN\,2013ha & 2013/11/06&  MCG +11-08-25 & 06:15:49.85 & 66:50:19.40& 0.0131 &LT &0.808 &- &- & -\\
    SN\,1987C & 1987/03/21& MCG+09-14-47 & 08:30:01.30 & 52:41:33.00& 0.0142 &LT &0.146 &- &0.632 & 0.926 \\
    SN\,2016ehw & 2016/07/20& MCG+12-08-47 & 08:36:37.60 & 73:35:03.70& 0.0120 &LT &0.041 &- &- &- \\
    ASASSN-15lf & 2015/06/15&  NGC 4108 & 12:06:45.56 & 67:09:24.00 & 0.0084&  LT&0.000 &- & 0.638 & 0.645 \\
    SN\,2012ab & 2012/01/31& A122248+0536 & 12:22:47.60 & 05:36:25.00& 0.0180 &LT &0.912 &- &- &- \\
    PS\,15aip & 2015/05/02&  KUG 1319+356 & 13:21:55.23 & 35:21:32.00 & 0.0195&LT &0.860 &- & -&- \\
    SN\,2006M & 2006/01/17& PGC 47137 & 13:27:19.76 & 31:47:14.50 & 0.0150&LT &0.575 &- & -& -\\
    SN\,1978G & 1978/11/24& IC 5201 & 22:20:48.30 & -46:01:22.00& 0.0031 &LCOGT\,2m & 0.000&- & -&- \\
    SN\,2006dn & 2006/07/05& UGC 12188 & 22:47:37.84 & 39:52:50.16& 0.0171 &LT &0.434 &- &0.059 &0.053 \\
\hline
\end{tabular}
\end{scriptsize}
\end{center}
\end{table*}

\subsection{NCR analysis results}

\begin{figure} 
\centering
  \includegraphics[width = \columnwidth]{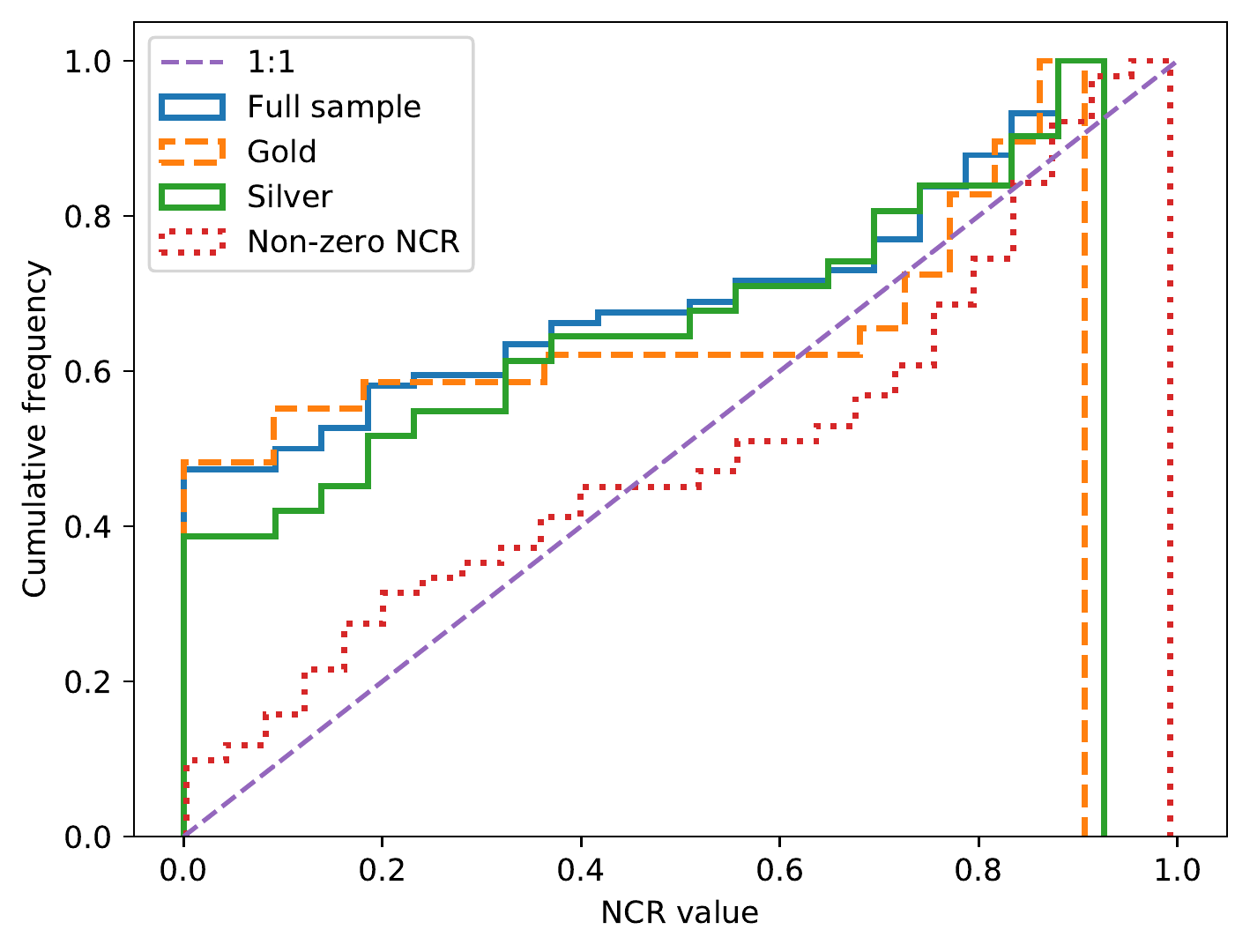}
  \caption{Cumulative distribution of the NCR values of the subsamples of our SNe\,IIn environments. The solid blue line is the full sample, the orange dashed line is the gold sample, the solid green line is the silver sample and the red dashed line is the non-zero NCR value sample. Also plotted is the 1:1 relationship (purple dashed line) which represents a population of progenitors which are strongly associated with star formation as traced by H$\alpha$ emission.}
  \label{NCR_results}
\end{figure}

 We show the results of our NCR analysis in \figureref{NCR_results}. We compare sub-samples of SN\,IIn NCR values to a 1:1 NCR value relationship, which represents a hypothetical population of transients that perfectly traces H$\alpha$ emission (and therefore, ongoing star formation). The full sample has a mean NCR value of 0.306\,$\pm$\,0.041 and the non-zero NCR subsample has a mean of 0.521\,$\pm$\,0.038. A mean NCR value of around 0.500 would suggest a population that follows the emission.

In \tableref{NCRKS} we present the AD p-values of different sub-samples of our NCR values when compared to each other and to a hypothetical population that perfectly follows H$\alpha$ emission. An AD test p-value $\leq0.05$ suggests that two populations differ significantly and are likely to be drawn from separate parent populations. We find that our full sample (blue line in \figureref{NCR_results}) of SNe\,IIn does not follow star-formation as traced by H$\alpha$ emission when compared with the hypothetical population that traces star-formation perfectly, with an AD p-value  $\sim$\,10$^{-6}$. We then split our full sample into the gold and silver spectral subcategories from \citetalias{Ransome_2021} and again these sub-samples likely do not follow the H$\alpha$ emission and are similar to each other and the full sample. However, when we split the sample and isolate the non-zero NCR SNe\,IIn  we find that the AD p-value (0.37) is consistent with the non-zero NCR sub-sample being drawn from the same population as the perfectly 1:1 relation. Therefore, the non-zero NCR SN\,IIn sub-sample is likely to follow the H$\alpha$ emission and therefore appears correlated with (on-going) star formation. 

The statistics so far suggest that we are observing multiple populations within the SN\,IIn class when considering NCR distributions. In order to test for multi-modality, we utilise HDT and the bimodality co-efficient. A HDT p-value $\leq0.05$ indicates the sample is significantly multi-modal and a HDT p-value of $0.05-0.10$ suggests multi-modality of marginal significance. The bimodality coefficient uses the skew and kurtosis of a distribution and bimodality coefficient of $>5/9$ suggests a bimodal distribution.

We have applied these tests to our sub-samples with the results presented in \tableref{bimodstats}. The HDT p-values indicate that the full sample is at least bimodal with HDT p-values in the order $\sim$\,10$^{-3}$. However and non-zero NCR sub-samples do not show significant levels of bimodality according to the HDT. None of our sub-samples have a bimodality coefficient $>5/9$, which suggests our NCR values do not show bimodality. However this does not rule out multi-modality as suggested by the HDT p-values for the full sample.

\subsection{Radial analysis results}

\begin{figure*}
\includegraphics[width = \columnwidth]{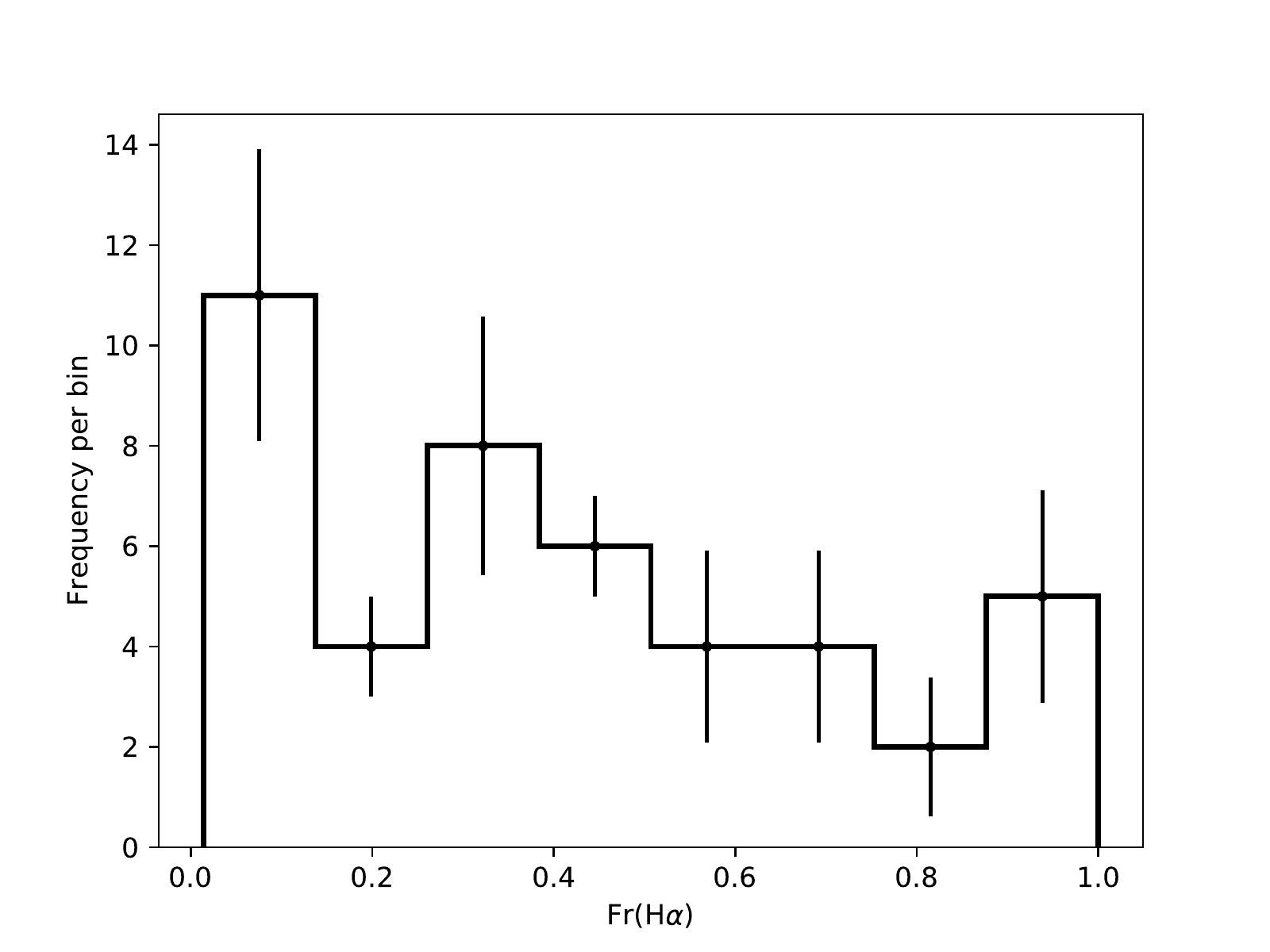}\hfill
\includegraphics[width = \columnwidth]{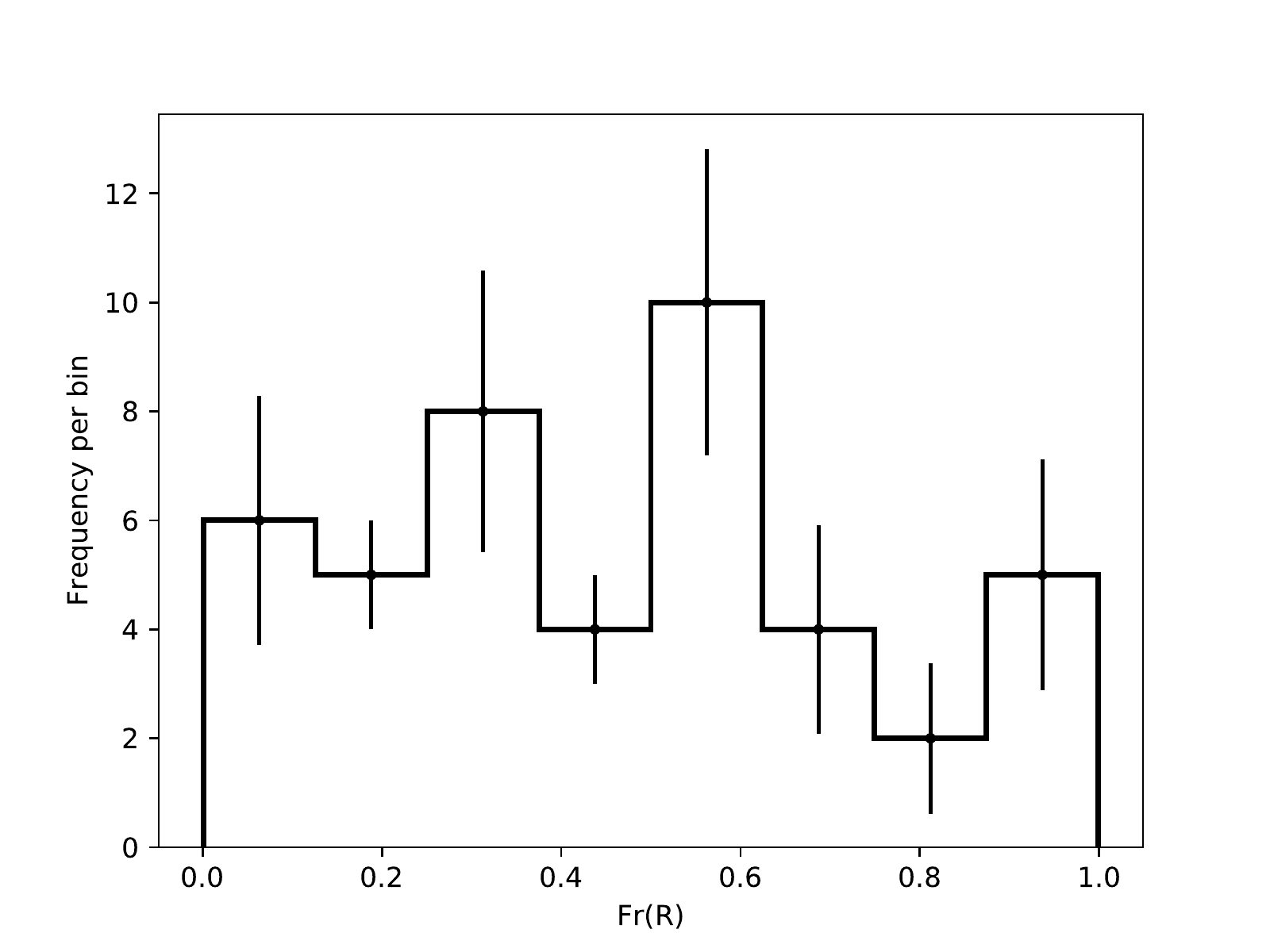}
\caption{Left: Fraction of H$\alpha$, Fr(H$\alpha$) flux contained within an ellipse that just encloses the SN pixel. Right: Fraction of $r'$-band flux contained within the ellipse that just enclosed the SN pixel. The y-axis on these plots is the number of SNe in each bin.\label{rad_ha}}
\end{figure*}

We present histograms showing the distributions of the fraction of $r'$-band or H$\alpha$ flux contained within the ellipse that just intersects the SN pixel. \figureref{rad_ha} shows the histogram for the distribution of the H$\alpha$ fraction, Fr(H$\alpha$), and $r'$-band faction, Fr(R). 

The sample of 45 SN\,IIn hosts used in the radial analysis is smaller than the full sample as this analysis requires hosts that are well defined such that a radial profile can be constructed. This will rule out hosts with semi-major axes of under $\sim$\,0.5'. The mean Fr(R) value is 0.454\,$\pm$\,0.040 and the mean Fr(H$\alpha$) values is 0.413\,$\pm$\,0.042. From the histograms in \figureref{rad_ha} we can see that there may be a hint of a peak at lower Fr(H$\alpha$) values which indicates centrally concentrated SNe but generally, the radial H$\alpha$ distribution is fairly evenly distributed. The Fr(R) distribution tells a similar story, the radial distribution is fairly evenly distributed. More data is needed in order to determine any possible trends here. This is consistent with the findings of \citet{hab14}.

In \tableref{bimodstats} we show the  the HDT p-values and the bimodality coefficients for our Fr(R) and Fr(H$\alpha$) value distributions. For both of these distributions, no bimodality was found by either the HDT or the bimodality coefficient.

\begin{table}
\caption{The multi-sample Anderson-Darling test results comparing the NCR values of our sub-samples of SNe\,IIn. These tests probe whether or not the sub-samples are drawn from the same parent population.\label{NCRKS}}
\begin{center}
\begin{tabular}{cc}
\hline 
     Sample & AD p-value\\
     \hline
     Full sample -- 1:1 & 2.7\,$\times$\,10$^{-8}$\\
     Gold -- 1:1 & 8.5\,$\times$\,10$^{-5}$\\
     Silver -- 1:1 & 1.0\,$\times$\,10$^{-3}$\\
     Non-zero NCR -- 1:1 & 0.37\\
     Gold -- Full sample &0.79 \\
     Silver -- Full sample &0.72\\
     Non-zero NCR -- Full sample & 3.6\,$\times$\,10$^{-6}$\\
     Gold -- Silver &0.27\\
     Gold -- Non-zero NCR & 7.0\,$\times$\,10$^{-5}$\\
     Silver -- Non-zero NCR &4.3\,$\times$\,10$^{-3}$\\
\hline
\end{tabular}
\end{center}
\end{table}

\begin{table}
\caption{The Hartigan's dip test and bimodality coefficients for our sub-samples of SNe\,IIn and the flux fractions.\label{bimodstats}}
\begin{center}
\begin{tabular}{ccc}
\hline
         Sample& \,HDT p-value\, & \,Bimodality coefficient\,   \\
     \hline
     Full sample & 7.9\,$\times$\,10$^{-3}$ & 0.23 \\
     Non-zero NCR &4.7\,$\times$\,10$^{-2}$ &0.19 \\
     Fr(H$\alpha$) & 0.87& 0.32\\
     Fr(R) &0.38 & 0.26\\
 \hline
\end{tabular}
\end{center}
\end{table}

\section{Discussion} \label{Discuss}

We have carried out the largest H$\alpha$ survey of nearby ($z<0.02$) SN\,IIn hosts, almost quadrupling the sample size from previous studies. The association of the environments of a sample of 77 nearby SNe\,IIn to H$\alpha$ emission and therefore ongoing star-formation was measured.

\subsection{Constraining progenitor systems} \label{lbv}

The relations in the NCR values presented in \sectionref{results} enable inferences to be made on the underlying stellar population. \citet{and12} investigated the association of a large sample of CCSNe with the host galaxy star formation using the NCR pixel statistics method and found a mass ladder in terms of progenitor mass and NCR value distribution.

 Some SNe\,Ic progenitors are WR stars \citep{Georgy_2009} and LBVs may be a preceding evolutionary stage to WR stars. It is known that at least some SNe\,IIn have LBV progenitors \citep[e.g, SN\,2005gl][]{Gal-Yam_2007}. One would therefore expect that SNe\,IIn would follow ongoing star-formation as traced by H$\alpha$ emission if the main progenitor path for SNe\,IIn were LBVs. However, this is not what we observe in our environmental analysis. We found that over 40\% of our sample had an NCR value of zero, indicating the SN pixel had \textbf{no} association with the H$\alpha$ emission. When we isolate the non-zero NCR values we find that when we compare the NCR values to a 1:1 relationship, we find that the non-zero NCR values follow this 1:1 distribution. The AD test p-values suggest these two samples are drawn from the same parent population. This may indicate that there are at least two populations of SNe\,IIn: a non-zero NCR group that follows H$\alpha$ emission and a population of zero NCR values. This may indicate that these SNe\,IIn have different progenitor stars. Instead, if we assume that LBVs are the progenitor system of the vast majority of SNe\,IIn (with the rest being made up of transients such as SNe\,Ia-CSM or ecSNe) then this may inform us about the evolution and environments of LBVs. Some LBVs may reside in areas outside of a star formation region in their hosts \citep{snt15}. \citeauthor{snt15} found that the neither the Galactic LBVs or LBVs in the LMC and SMC were well associated with O-type star clusters. Further to this, \citeauthor{snt15} note that in the LMC, the LBVs were more isolated than the Galactic or SMC LBVs and were more isolated than the known WR stars in the LMC. Those authors suggest that LBVs may not be single stars but evolve in binaries \citep[as many O-type stars do,][]{Gies_1987, Garcia_2001, Evans_2006}. Furthermore, LBVs that are isolated and apparently in single star systems may have evolved in a multiple star system and then either the companion star has exploded as a SN or the LBV got kicked out of the system due to interactions with nearby stars and became a ``runaway'' star. This may cause a spin-up of the LBV, which is consistent with the models of \citet{Groh_2013} where the LBVs in their models exploded as SNe\,IIn when a rotational element was added. \citet{Aghakhanloo_2017} implemented models of LBV isolation and found that observed isolation of some LBVs is consistent with binary evolution. \citet{Smith_2020} present a study on the LBV candidate, MCA-1D (also known as UIT003) in the outskirts of M33. \citeauthor{Smith_2020} find that similarly to some observed SN impostors, MCA-1D had an outburst in 2010 that had a similar light-curve to other LBV outbursts. This LBV candidate is associated with a small cluster but is on the outskirts of the galaxy and the environment is similar to that of the impostor turned (possible) SN\,IIn, SN\,2009ip (see \sectionref{09ip}). \citet{Humphreys_2016} found that the velocities of seven M31 LBVs and seven M33 LBVs were inconsistent with a runaway star scenario. 

Generally, we have found that the progenitors of SNe\,IIn are longer lived than their host star forming regions which is indicated by the low average NCR values. We also found that the non-zero NCR subsample follows the H$\alpha$ emission, indicating that the progenitors in the non-zero NCR subsample resided in their star formation regions. Therefore we observe a young population and an older population. These two populations suggest that the SNe\,IIn do not have a single progenitor path. There is potential degeneracy in the environments of LBVs or LBV-like objects which may contaminate the non-zero NCR sample. Perhaps the contrast we see in environments can not simply be pinned to two sets of separate progenitors, with a higher mass component following ongoing star-formation and another population with lower mass progenitors that are less or unassociated with star formation. This would be consistent with the HDT and bimodality coefficient test results. These open questions prevent a clean distinction between the progenitor environments and masses as LBVs may be isolated and in OB star associations, have an unstable LBV classification or some may even be be formed by lower mass stars merging.
 
 Furthermore, when we split the sample into the gold and silver spectral categories \citepalias{Ransome_2021}, we find that the gold and silver SNe\,IIn are drawn from the same parent population, as indicated by the AD test p-values. This indicates that the objects in the two categories may similar progenitor types and there is there is unlikely to be a spatial difference between the gold and silver groups. Therefore this indicates that many of the silver SNe\,IIn may be promoted to a fully fledged gold SN\,IIn if follow-up observations were taken as suggested by \citetalias{Ransome_2021}. 

The inference that transients strongly associated with star formation as traced by H$\alpha$ emission are more likely have very massive progenitors is based upon the assumption that such transients have young, massive progenitors. H$\alpha$ emission has a characteristic timescale of $<16$\,Myr \citep{Haydon_2018, Haydon_2020} so one would expect that SNe appearing in these regions \citep[correlated with \ion{H}{ii} regions][]{Kuncarayakti_2013} are young and massive and therefore short-lived. They have not had enough time to drift from the star-forming region before their death.

As discussed here and in \sectionref{sec:intro} there are multiple possible progenitor paths for SNe\,IIn which may account for our observed multiple populations in NCR value. 

\subsubsection{Type Ia-CSM and SN\,2006gy} \label{iacsm}

As we have mentioned in \sectionref{sec:lowmassprog}, another scenario that may contribute to the SN\,IIn phenomenon are SNe\,Ia-CSM. SN\,Ia systems are old and will outlive their parent star formation region. \citet{hab14} compares the NCR values of their sample of SNe\,IIn with other classes of SNe, including 98 SNe\,Ia. Just under 60\% of the SNe\,Ia in that sample had a zero NCR value and most of the remainder were of lower NCR values, the slope on the cumulative frequency plot flattens out after around an NCR value of 0.600. The mean NCR value of the SNe\,Ia in \citeauthor{hab14} is 0.157. SNe\,Ia have mostly low or zero NCR values, therefore, SNe\,Ia-CSM may account for some of the zero and low NCR value SNe\,IIn in our sample.

A distinction between core-collapse SNe\,IIn and SNe\,Ia-CSM would not be directly picked up by the selection criteria set out by \citetalias{Ransome_2021} as the classification system uses only the H$\alpha$ line profile and would require the transient to be a recognised SN\,Ia-CSM. In this study we have two possible SN\,Ia-CSM, SN\,2006gy and SN\,2008J for which we have continuum subtracted H$\alpha$ data. However, we find that SN\,2008J has an NCR value of 0.808 and SN\,2006gy has a very high NCR value of 0.907, indicating these transients are in a strong, active star forming region. Alternatively to the thermonuclear origin of SN\,2006gy, this transient may have a massive progenitor. For example, \citet{Smith_2010} suggest that the $\sim$\,20\,M$_\odot$ of CSM required a very high mass progenitor with mass $\sim$\,100\,M$_\odot$.  As most SNe\,Ia are not associated with H$\alpha$ emission \citep{and12, hab14} this is unusual but one of the SNe\,Ia in the previous studies was in a region with an NCR value of over 0.900. As the thermonuclear origin of some of these transients becomes apparent at later times such as with SN\,2006gy, it is possible that some of the transients in our sample would be made up of hitherto unknown SNe\,Ia-CSM. These could therefore make up some of the zero and low NCR value SNe\,IIn, assuming that SNe\,Ia-CSM are in similar environments to SNe\,Ia. However, \citet{Silverman_2013} investigated 16 SNe\,Ia-CSM and found that all of them were found in spiral galaxies, suggesting that SNe\,Ia-CSM generally occur in younger populations when regular SNe\,Ia occur in all Hubble types \citep[with some SN\,Ia subtypes preferring elliptical galaxies, see][]{Hakobyan_2020}. Therefore, SNe\,Ia-CSM may, on average, have higher NCR values than regular SNe\,Ia \citep[for example those in ][]{and12, hab14}. As the CSM may be created by a super-AGB companion, it is possible that SNe\,Ia-CSM may have a similar average NCR value to SNe\,IIP or ecSNe as discussed in \sectionref{ecsn}. SNe\,Ia-CSM are rare \citep{Graham_2019} and for example, out of 127 SNe\,Ia observed by ZTF, only one was observed to be a SN\,Ia-CSM \citep{Yao_2019}. This rate may be overstated as SNe\,Ia-CSM are more luminous than SNe\,Ia, thus more easily observed. Due to the rarity of CSM interaction in SNe\,Ia, they likely do not make up a large ``hidden'' proportion of our SNe\,IIn sample, despite being the most numerous observed SN class \citep{Li_2011}. On the other hand, a number of SNe\,Ia show late time CSM interaction, such as SN\,2015cp that showed CSM interaction 664 days post-explosion \citep{Graham_2019_15cp} so some SNe\,Ia may be unrecognised as SNe\,Ia-CSM if they lack late time observations.

\subsubsection{Electron-capture supernovae and SN\,2008S} \label{ecsn}

\citet{Cai_2021} examine the intermediate luminous red transient (ILRT) phenomenon and the possible connection to ecSNe. It was found that all five of these ILRTs resembled SN\,2008S (which as well as being a prototypical impostor/transitional impostor to SN\,IIn is also considered a prototypical ILRT). The spectra show SN\,IIn-like narrow components to the H$\alpha$ profiles so would be classified as SNe\,IIn by \citetalias{Ransome_2021}. The NCR value of SN\,2008S was zero and ILRTs could make up a zero or low NCR value population of SNe\,IIn. SN\,2008S exploded in a dusty environment. \citet{hab14} cross-referenced 2MASS data of the hosts in their sample and found there was no underlying H$\alpha$ region that was being obscured by a dusty environment. 

Another example of a possible SN\,IIn with ecSN origin that is in our sample is SN\,2015bf. This object has an NCR value of 0.657, which indicates a moderately strong region of ongoing star-formation. This contrasts with the zero NCR value of SN\,2008S. \citet{Lin_2021} present a study on SN\,2015bf, they find that the spectrum evolves into a more standard SN\,II spectrum. SN\,2015bf had a fairly high peak luminosity of -18 but the light curve then started to decay rapidly with a similar light curve morphology to other fast-declining SN\,II, including the ecSN candidate, SN\,2018zd. Those authors argue that the brief CSM interaction seen in the spectral evolution points towards the CSM being fairly confined. This was interpreted as a violent mass loss episode shortly before the SN explosion. As ecSNe are expected to have progenitors on the low end of the progenitor mass range for CCSNe, they would be lower down the NCR value mass ladder. SN\,2015bf eventually evolved as a SN\,II and \citet{and12} found that around 30\% of SNe\,II had zero NCR values. Furthermore around 75\% of these SNe had an NCR value under 0.500, so SNe\,IIn from ecSNe may make up some of our lower or zero NCR population.

\subsubsection{Mass-loss continuum} \label{masslosscont}

SNe from lower mass progenitors such as ecSNe exemplify that mass loss sufficient to create enough CSM to produce the SN\,IIn phenomenon can be experienced by progenitors on the lower mass range for CCSNe. SNe\,IIn from lower mass progenitors such as RSGs may be classified as SNe\,IIn-P \citep[SNe\,IIn with SN\,IIP-like light curves, e.g. SN\,2013fs and PTF\,11iqb][]{Bullivant18, Smith_2015}  or SNe\,IIn-L \citep[SNe\,IIn with SN\,IIL-like light curves, e.g. SN\,2013fr and SN\,1998S][]{Bullivant18, tad15}. 

In this work we find five examples of possible ecSNe: SN\,2011ht \citep{Roming_2011}, SN\,1994W \citep{Sollerman_1998}, SN\,2009kn \citep{Kankare_2012}, SN\,2008S \citep{Botticella_2009} and SN\,2015bf \citep{Lin_2021}. All but one (SN\,2015bf) have an NCR value of zero. This may be consistent with these objects having lower mass progenitors. Furthermore, the NCR distribution of standard SN\,II in \citet{and12} finds half the SNe in zero NCR regions. 

\citet{Boian_2020} model early time spectra of SNe that exhibit early time CSM interaction. \citeauthor{Boian_2020} use these models to explore the possible progenitors of 17 SNe with early time interaction. A wide range of SN parameters were calculated for this sample along with finding that there was increased mass loss immediately preceding the SN explosions. This early time interaction is not exclusive to the classic SN\,IIn class however, as, for example flash spectrum SNe also exhibit this behaviour and may have RSG progenitors \citep{Khazov_2016, Dessart_2017, Kochanek_2019}. Using the classification criteria in \citetalias{Ransome_2021}, it is possible that some of the silver class SNe\,IIn that have a limited number of spectra are actually flash spectrum SNe but further data is needed to demote these objects from silver SNe\,IIn.

\subsection{SN 2005ip and long-lasting SNe IIn} \label{longlast}

A sub-category of SNe\,IIn are the long lasting SNe\,IIn. Examples of this phenomenon include SN\,1988Z \citet{Turrato_1993, Chugai_1994}, SN\,1995G \citep{Pastorello_2002, Chugai_2003}, SN\,2005ip \citep{Stritzinger_2012, hab14, Smith_2016b}, SN\,2015da \citep{Tartaglia20}, and KISS15s \citep{Kokubo_2019}. 

SN\,2005ip is perhaps the most well studied long-lasting SN\,IIn. \citet{hab14} found that SN\,2005ip was still the brightest H$\alpha$ source in its host, three years post explosion. \citet{Fox_2020} present a study on SN\,2005ip, including (very) late time photometry and find that SN\,2005ip had only just started to decline in 2015. We find that SN\,2005ip is no longer the strongest H$\alpha$ source in the host with an NCR value of 0.866. However we do not include SN\,2005ip in our analysis as we do not know whether the transient has dimmed to quiescent levels. Another example of a long-lasting SN\,IIn in our sample is SN\,2015da. \citet{Tartaglia20} found that SN\,2015da was still slowly declining four years post-explosion. Those authors suggest that the slow decline was due to ongoing CSM interaction with a CSM mass of $\sim$\,8\,M$_\odot$ with an extreme pre-explosion mass loss rate of $\sim$\,0.6\,M$_\odot$\,yr$^{-1}$. SN\,2015da has a NCR value of 0.997, indicating that SN\,2015da is in one of the strongest H$\alpha$ emission regions in the host. This very high NCR value may be skewed by the ongoing interaction and is therefore not included in the NCR analysis. We do not find any additional examples of long-lasting SNe\,IIn in our observations.

\subsection{SN\ 2009ip and precursors} \label{09ip}

\citet{Ofek_2014, Strotjohann_2020} found that many SNe\,IIn suffer from precursor explosions which may be LBV-like great eruptions. Notable examples include SN\,2009ip \citep{Foley_2011, Mauerhan13a, Pastorello_2013}, SN\,2011ht \citep{Roming_2011}, and SN\,2015bh \citep{Boian_2018, Thone_2017}. 

 We find that SN\,2009ip is a very isolated transient, with an NCR value of zero and the similar transients, SN\,2015bh and SN\,2011ht were also zero NCR SNe\,IIn. If these precursor ``impostor''  events are LBV eruptions, these could be further examples of LBV isolation, possibly showing the effects of binary evolution. 
 
 In some cases there is little pre-explosion data. \citet{Groh_2013b} suggest that if the archival images of the progenitor of SN\,2005gl were taken at the time the progenitor was suffering a pre-SN\,IIn outburst, then the mass estimates for the progenitor may be overestimated. If this were the case, the progenitor may have a mass of around 20-25M$_{\odot}$.

SN impostors are an important consideration, and we can not rule out there being contamination in our sample (especially in the case of the silver SNe\,IIn). An environmental study of this nature, with a sample of SN impostors is difficult due to selection effects involved. Impostors are generally dim events so we may be biased towards exceptional, bright objects, especially if the objects are superimposed on a bright \ion{H}{ii} region. A study of the environments of SN impostors is therefore beyond the scope of this paper.

\subsection{Radial distributions}  \label{raddist}

We adopted the radial analysis used in \citet{Anderson_2009} and later in other studies \citep{Habergham_2012,hab14} in which the spatial distribution of our transients is compared with respect to younger stellar populations (traced by the H$\alpha$  emission) and an older population (traced by the $r'$-band emission). Our findings were consistent with previous studies \citep[e.g., in][]{hab14}. In terms of the H$\alpha$ emission, spatially, there may be a weak central concentration of SNe\,IIn shown by a possible peak in lower Fr(H$\alpha$) values. There may be a peak in intermediate Fr(R) values and a slight decline in frequency at higher Fr(R) values, indicating a dearth of SNe at the periphery of the host $r'$-band flux. The peak at intermediate values in the Fr(R) distribution may correspond with the possible central concentration in the Fr(H$\alpha$) as there tends to be less H$\alpha$ emission in the central regions of galaxies so a low Fr(H$\alpha$) value may be further into the disc and in the mid-values of Fr(R).

We did not find a central excess in the radial distributions of SNe\,IIn. This does not follow the distributions seen for high mass (and high mass loss) progenitors such as those of SNe\,Ic. SNe\,Ic tend to be centrally located \citep{vandenBergh_1997, Habergham_2012} and as the radial distribution is a proxy for metallicity, are in higher metallicity environments.

We did not include SN impostors in this work. When \citet{hab14} included impostors in their study, they found that the Fr(H$\alpha$) distribution had a peak at high values. This indicated that the impostors tended to be at the periphery of the host H$\alpha$ emission. If many of the impostors are great eruptions of LBVs, then this would be consistent with the apparent isolation of LBVs we discussed in \sectionref{lbv}. The authors also found that there was a dearth of SNe in the central regions with very central SNe with respect to Fr(R), however do find central transients. Spatially there are a number of very central SNe\,IIn in our sample and \citetalias{Ransome_2021} verified that these transients were not AGN.

\subsection{Selection effects and detection sensitivity}

We collated our sample from online databases which compile transients from many sources which include surveys that may be targeted or untargeted. Any bias inherent in these surveys will persist through to our target list. For example, large surveys such as the Zwicky Transient Survey \citep[ZTF, ][]{ztf} may introduce bias when transients are selected for spectroscopic followup. 

The NCR method may lead to missing low level H$\alpha$ emission. As the zero NCR population are the SNe where the cumulative sum is under zero, there may be a skew towards the positive pixel values and negative values, indicating that the low level emission is being missed. Therefore, we may be overestimating the zero NCR population. This could be remedied with larger telescopes with capabilities to perform deeper observations.

Another potential source of bias in our work is that we exclude hosts with smaller angular diameters from the radial distribution analysis.

\subsection{Summary}

We have presented the results of the largest environmental study of SN\,IIn hosts to date with 77 SNe\,IIn with most having a strong spectral classification. Our conclusions can be summarised as:

\begin{enumerate}
    \item We found that as a whole, SNe\,IIn do not follow star-formation as traced by H$\alpha$ emission. 
    \item We find that around \textbf{40\%} of SNe\,IIn are not associated with any SN\,IIn emission as calculated by NCR.
    \item The non-zero NCR population is consistent with the hypothetical star-formation following population.
    \item Our findings suggest there may be multiple progenitor routes to SNe\,IIn (e.g. ecSNe or SNe\,Ia-CSM). We do not see bimodality in our NCR distributions but we see multimodality in the full sample and non-zero NCR value subsample.
    \item There are no significant differences in the NCR distributions of the gold and silver classes. This suggests that many of the silver SNe\,IIn may be promoted to gold SNe\,IIn given more spectra.
    \item The radial distributions of SNe\,IIn in terms of the $r'$-band and H$\alpha$ emission is even. However we do note there are more centrally located SNe\,IIn than previous studies. While there is no central excess found in the distribution, there may be some SNe\,IIn progenitors that are similar to the progenitors of the massive, centrally concentrated SNe\,Ic. 
\end{enumerate}

 Future surveys will provide a huge amount of data and transient discoveries. Surveys such as the ZTF and the Legacy Survey of Space and Time (LSST) at the Vera C. Rubin Observatory \citep{LSST} will provide a wealth of SN\,IIn candidates and will allow much larger samples to be used for constraining possible SN\,IIn progenitors.

\section*{Acknowledgements}

 We would like to thank the anonymous referee for their helpful and insightful comments. The Liverpool Telescope is operated on the island of La Palma by Liverpool John Moores University in the Spanish Observatorio del Roque de los Muchachos of the Instituto de Astrofisica de Canarias with financial support from the UK Science and Technology Facilities Council (STFC). This work makes use of observations from the Las Cumbres Observatory global telescope network. C.L.R.\ acknowledges a PhD studentship from STFC. S.M.H.-M.\ and M.J.D.\ acknowledge partial funding from STFC. The Isaac Newton Telescope and its service programme are operated on the island of La Palma by the Isaac Newton Group of Telescopes in the Spanish Observatorio del Roque de los Muchachos of the Instituto de Astrofísica de Canarias.

\section*{Data Availability}

The data used in this work will be shared upon reasonable request to the author.





\bibliographystyle{mnras}
\bibliography{example} 




\appendix


\bsp	
\label{lastpage}
\end{document}